\pdfoutput=1
\documentclass[11pt,twoside,a4paper,cmspaper,final,collab]{cms-tdr}
\def\svnVersion{9562ed0}\def\svnDate{2026/08/15}\def\cmsCernNoTag{CERN-EP-2026-017}\def\cmsCernDate{\today}\def\cmsMessage{Published in Physical Review Letters as \href{http://dx.doi.org/10.1103/g49y-8cjl}{\doi{10.1103/g49y-8cjl}.}}
\begin{document}\cmsNoteHeader{HIN-25-012}

\newcommand{\PbPb}{\ensuremath{\text{PbPb}}\xspace}
\newcommand{\pp}{\ensuremath{\text{pp}}\xspace}
\newcommand{\NN}{\ensuremath{\text{NN}}\xspace}
\newcommand{\roots}{\ensuremath{\sqrt{s}}\xspace}
\newcommand{\HYBRID}{H\textsc{ybrid}\xspace}
\newcommand{\pth}{\ensuremath{p_{\mathrm{T}}^{\text{ch}}}\xspace}
\newcommand{\jtetaone}{\ensuremath{\eta^{\text{jet}_1}}\xspace}
\newcommand{\jtetatwo}{\ensuremath{\eta^{\text{jet}_2}}\xspace}
\newcommand{\deta}{\ensuremath{\Delta\eta}\xspace}
\newcommand{\dphi}{\ensuremath{\Delta\varphi}\xspace}
\newcommand{\dphijt}{\ensuremath{\dphi^{\text{jet}_{1},\text{jet}_{2}} > 5\pi/6}\xspace}
\newcommand{\detaldtrk}{\ensuremath{\deta^{\text{ch}, \; \text{jet}_{1}}}\xspace}
\newcommand{\dphildtrk}{\ensuremath{\dphi^{\text{ch}, \; \text{jet}_{1}}}\xspace}
\newcommand{\Rsym}{\ensuremath{\text{R}^{\text{sym}}}\xspace}
\newcommand{\Rasym}{\ensuremath{\text{R}^{\text{asym}}}\xspace}
\newcommand{\detajtonetwo}{\ensuremath{\abs{\deta^{\text{jet}_{1},\text{jet}_{2}}}}\xspace}
\ifthenelse{\boolean{cms@external}}
{
  \renewcommand{\appendixname}{End Matter}
  \newcommand{\appname}[1]{\appendixname~\ref{#1}}
}
{
  \newcommand{\appname}[1]{\appendixname~\ref{#1}}
}

\cmsNoteHeader{HIN-25-012}

\title{Observation of the jet diffusion wake using dijets in heavy ion collisions}
             
\abstract{Energetic quarks and gluons traversing a hot and dense quark-gluon plasma deposit energy and momentum into the medium before hadronizing to collimated sprays of particles, known as jets. This energy-momentum deposition is expected to produce medium responses, collectively known as jet wakes, with ``diffusion wake'' denoting a depletion of particles in the direction opposite to the propagating jet. The diffusion wake is studied by comparing dijet-hadron correlations measured in lead-lead (\PbPb) and proton-proton (\pp) collisions. The analysis uses \PbPb and \pp data recorded at a nucleon-nucleon (\NN) center-of-mass energy $\sqrtsNN = 5.02\TeV$ with the CMS detector at the CERN LHC. By exploring how the dijet-hadron correlation distributions differ for various pseudorapidity separations of the two jets in the dijet, the presence of a jet diffusion wake is firmly established. The diffusion wake signal has a significance greater than 5 standard deviations for charged particles in the transverse momentum range $1 < \pt < 2\GeV$. The measurements are compared with model predictions with and without jet wake effects, providing new insights into quark-gluon plasma properties and the formation of jet-induced wakes.}
\hypersetup{
pdfauthor={CMS Collaboration},
pdftitle={Observation of the jet diffusion wake using dijets in heavy ion collisions},
pdfsubject={CMS},
pdfkeywords={CMS, physics, heavy-ion collision, jets, quark-gluon plasma, diffusion wake}}

\maketitle
Jets are collimated streams of particles resulting from hard-scattered quarks and gluons, collectively known as partons, and are effective probes of the quark-gluon plasma (QGP)~\cite{Busza:2018rrf}. The QGP, a hot form of matter consisting of deconfined partons, is created in ultrarelativistic heavy ion collisions, and its study enables experimental tests of quantum chromodynamics. The opacity of QGP to the propagation of partonic probes causes substantial energy losses for the partons that traverse the hot medium. This opacity manifests itself in various experimental observables, such as suppression of high transverse momentum (\pt) hadrons and jets, as well as modifications of jet showers~\cite{CMS:2024krd, Cunqueiro:2021wls, Apolinario:2022vzg}. Collectively referred to as jet quenching, these phenomena were first observed at the BNL RHIC~\cite{PHENIX:2001hpc, STAR:2002ggv} and later at the CERN LHC~\cite{Chatrchyan:2011sx, ATLAS:2010isq, ALICE:2015mdb}.

Quenching and its impact on jets have been studied extensively~\cite{CMS:2024krd, Cunqueiro:2021wls, Apolinario:2022vzg}. However, separating jet modifications from jet-induced medium response remains challenging. It is theoretically clear that both types of modification coexist for jets in a QGP medium. Disentangling them will improve the understanding and modeling of medium evolution and dynamics.

Jet-medium interactions were predicted to create a Mach-cone-like effect in the QGP~\cite{Casalderrey-Solana:2004fdk, Stoecker:2004qu, Ruppert:2005uz, Gubser:2007ga, Neufeld:2008fi, Qin:2009uh, Li:2010ts, Bouras:2012mh, Ayala:2016pvm, Yan:2017rku, Casalderrey-Solana:2020rsj}. Experimental observation of this effect remains elusive because the Mach cone signal overlaps with medium-induced jet-shower modifications along the jet propagation direction. A related effect producing a depletion of particles in the opposite direction, known as a ``diffusion wake,'' is also expected~\cite{Gubser:2007ni, Betz:2008ka, Yang:2021qtl, Yang:2022nei, ATLAS:2024prm, CMS:2024fli}. The first evidence of a diffusion wake using \PZ-boson-hadron correlations was reported by the CMS experiment~\cite{CMS:2024fli}. Similarly, the ATLAS experiment searched for the diffusion wake in photon-jet events but found no significant evidence within uncertainties~\cite{ATLAS:2024prm}. 

Recent theoretical developments~\cite{Pablos:2019ngg, Yang:2025dqu} proposed the use of dijet events, where the two jets are back-to-back in azimuthal angle ($\varphi$), but separated in pseudorapidity ($\eta$), to spatially separate the jet modification and diffusion wake. Specifically, the hadron distributions associated with dijets of sizeable $\eta$ separation compared to those without separation, are predicted to provide a robust signal of the diffusion wake.

This Letter presents the first measurement of the jet diffusion wake using dijet events. A sample of lead-lead (\PbPb) collisions at a center-of-mass energy per nucleon pair (\NN) of $\sqrtsNN = 5.02\TeV$ is used. A reference measurement is derived from proton-proton (\pp) collisions at the same energy. Data samples, corresponding to integrated luminosities of 0.66\nbinv for \PbPb collisions and 299\pbinv for \pp collisions~\cite{CMS-PAS-LUM-19-001, CMS-PAS-LUM-20-002}, were collected with the CMS detector at the LHC during 2018 and 2017, respectively. Dijets are defined as the two highest \pt back-to-back jets, labeled ``$\text{jet}_{1}$'' (leading) and ``$\text{jet}_{2}$'' (subleading). The back-to-back requirement is that the azimuthal angle separation of the jets satisfies \dphijt. For samples requiring different $\eta$ separations between the two jets in the dijet, the distributions of charged particles relative to the leading jet axis direction are studied as functions of \detaldtrk and \dphildtrk, where the superscript ``ch'' denotes charged particles. To isolate the diffusion wake signal, the correlation distributions obtained for dijets with a small $\eta$ gap are subtracted from those with large gaps. The resulting distributions from \PbPb events are compared to those measured using \pp collisions. The results are presented differentially in charged-particle \pth and \PbPb collision centrality, defined as the percentile of the total inelastic hadronic cross section (with 0\% corresponding to the most central collisions). The diffusion wake signal, seen as a depletion of the particle yield in the underlying-event (UE) distribution, is extracted and compared to theoretical models with and without medium effects. The tabulated results are provided in the HEPData record for this analysis~\cite{hepdata}.

A complete description of the CMS apparatus is reported in Refs.~\cite{CMSexperiment,CMS:2023gfb}. A ``particle-flow'' algorithm~\cite{CMS:2017yfk} aims to reconstruct all individual particles in an event, combining information provided by the all-silicon tracker, the crystal electromagnetic and brass-scintillator hadron calorimeters, and muon detectors. The forward hadron (HF) calorimeter consists of two halves located 11.2\unit{m} from the interaction region, one on each end, and provides coverage in the range $3.0 < \abs{\eta} < 5.2$.

The \pp and \PbPb samples were selected with a calorimeter-based trigger that uses the anti-\kt jet clustering algorithm with a distance parameter of $R=0.4$~\cite{Cacciari:2008gp}. This trigger is fully efficient for events with reconstructed jets of $\pt>130\GeV$. The \PbPb and \pp events selected for further analysis must have at least one primary interaction vertex containing at least two tracks~\cite{CMS:2014pgm} and located within 15\unit{cm} of the nominal interaction point along the beam axis. Background events from beam-gas interactions are removed off-line using the criteria described in Ref.~\cite{Khachatryan:2016odn} for both \PbPb and \pp collisions.

Each \PbPb event must have at least two detector elements (towers) in each HF detector that contain energy deposits exceeding 4\GeV per tower to remove contamination from ultraperipheral collisions~\cite{Baltz:2007kq}. Collision centrality is determined based on the total sum of transverse energy (defined as $\et=E/\cosh\eta$, where $E$ is the measured energy) detected in the HF region covering $3.0<\abs{\eta}<5.2$~\cite{Chatrchyan:2011sx}. Events in \PbPb collisions are divided into three centrality intervals: 0--30\% (most central), 30--50\%, and 50--80\% (most peripheral).

Simulated Monte Carlo (MC) events are used to correct for detector effects. The \PYTHIA 8.226~\cite{pythia82} event generator, configured with the CP5 tune~\cite{pythiaCP5tune} and NNPDF3.1 parton distribution functions at next-to-next-to-leading order~\cite{nnpdf31}, models the hard scattering processes, parton showering, and hadronization that produce jets of interest. These events are also embedded into simulated minimum bias \PbPb collisions, generated using \HYDJET 1.9~\cite{Lokhtin:2005px} to represent the \PbPb UE. The simulated sample, combining \PYTHIA hard interactions and \HYDJET underlying events, is called {\PYTHIA}+\HYDJET. The CMS detector response is simulated using the \GEANTfour~\cite{GEANT4} toolkit. A centrality-based reweighting is applied to the {\PYTHIA}+\HYDJET sample to match the centrality distribution of the jet-triggered \PbPb data. Additionally, a reweighting procedure is performed to match the distribution of the primary vertex position between the MC simulations and data.

Jet reconstruction is performed by the anti-\kt jet algorithm with a distance parameter of $R=0.4$, as implemented in the \FASTJET framework~\cite{Cacciari:2011ma}, using particles reconstructed by the particle-flow algorithm~\cite{CMS:2018jrd,CMS:2016lmd,CMS:2019ctu}. The constituent subtraction method~\cite{constituentSubtraction} is used to remove UE contributions. The jet axis direction in $\eta$--$\varphi$ is determined for each jet by the vector sum of the four-momenta of its constituent particles. Jet energy corrections are derived from MC simulations to make the average reconstructed jet energy identical to that of generated-level jets. \textit{In situ} measurements are used to correct for residual differences between data and simulation. 

After jet reconstruction and energy corrections, leading jets are required to satisfy $\pt^{\text{jet}_1} > 130\GeV$ and $\abs{\jtetaone} < 1.0$. Subleading jets must fulfill $\pt^{\text{jet}_2} > 50\GeV$ and $\abs{\jtetatwo} < 2.0$, in addition to the back-to-back requirement \dphijt. This kinematic selection ensures full trigger efficiency and optimal jet energy resolution (JER). The dijet sample is divided based on the $\eta$ separation of leading and subleading jets, with $\detajtonetwo < 0.5$ representing small gaps and $\detajtonetwo \in (0.5, 1.0)$, $\detajtonetwo \in (1.0, 1.5)$, and $\detajtonetwo \in (1.5, 2.0)$ representing large gaps.

The reconstruction of charged-particle tracks is described in Ref.~\cite{CMS:2014pgm}. Tracks with $1<\pth<2$ and $2<\pth<4\GeV$ within $\abs{\eta^{\text{ch}}}<2.4$ are used, and only high-purity tracks~\cite{CMS:2014pgm} with relative \pth uncertainty $\delta\pth/\pth$ below 10\% are considered. The distance of closest approach to the primary vertex in both the $x$--$y$ plane and the $z$ direction must have a significance below 3 standard deviations to reduce secondary-track contributions. In \PbPb collisions, tracks must additionally have ${>}$10 hits in the tracker layers and a $\chi^2$ per degree of freedom per silicon-detector layer below 0.18. 

Correlations between the tracks and the leading jet direction are constructed following Refs.~\cite{CMS:2016qnj,CMS:2016cvr,CMS:2018zze,CMS:2020geg,CMS:2021nhn}. Distributions are formed in (\detaldtrk, \dphildtrk) space for both \pth bins. Following Ref.~\cite{Yang:2025dqu}, an $\eta$ ordering of $\jtetaone > \jtetatwo$ is enforced; a reflection of the correlation distribution is made by reversing the sign of \detaldtrk when this condition is not satisfied. The correlated yields are normalized by the number of dijet pairs and corrected for tracking efficiency and misreconstruction, following the procedure outlined in Ref.~\cite{Khachatryan:2016odn}. Pair-acceptance effects due to limited detector acceptance are corrected using a mixed-event method~\cite{CMS:2013jlh,CMS:2016qnj,CMS:2016cvr,CMS:2018zze,CMS:2020geg,CMS:2021nhn}, in which the leading jet from one event is paired with charged particles from 30 other dijet events selected randomly. To ensure proper acceptance matching, mixing is performed only between events with similar primary vertex positions along the beam axis (within 0.5\unit{cm}) and collision centrality (within 0.5\%).

Additional corrections typically applied to account for jet position resolution and selection biases (favoring jets with harder constituents or those affected by upward \pt fluctuations) are omitted, as they cancel when subtracting correlation distributions with large and small $\eta$ gaps and do not affect the charged-particle densities in the UE, which is the focus of this study. The jet selections are made without corrections for detector resolution, as the observable is insensitive to jet energy resolution effects.

Examples of the final correlation distributions for \pp and 0--30\% \PbPb collisions are shown in \appname{app:2D_jetTrackCorr}. Two peaks are observed in the resulting correlations, as shown in Fig.~\ref{fig:2Dresults}. The near-side ($\dphildtrk \approx 0$) peak is associated with leading-jet particles, while the away-side ($\dphildtrk \approx \pi$) peak corresponds to the subleading jet. Because $\jtetaone > \jtetatwo$, the away-side jet peak and the associated diffusion wake would always appear in the negative $\detaldtrk$ region. For small dijet $\eta$ gaps, the correlation distribution (denoted as \Rsym) is roughly symmetric in \detaldtrk, with both peaks located around $\detaldtrk \approx 0$. Consequently, the diffusion wakes are hidden by the jet peaks, and no wake signal is visible. In contrast, for large $\eta$ gaps, the away-side peak and its diffusion wake are displaced toward negative \detaldtrk. The subleading-jet diffusion wake then manifests as a dip in the near-side correlation, while the leading-jet diffusion wake appears as a dip on the away side in the asymmetric distribution, \Rasym. 

Near-side ($\abs{\dphildtrk} < \pi/2$) projections of the two-dimensional correlations onto \detaldtrk, per radian, are used to isolate the diffusion wake signal. The long-range uncorrelated background is estimated from the positive \detaldtrk region ($1.5 < \detaldtrk < 2.5$) similar to previous jet--track correlation studies~\cite{CMS:2016qnj,CMS:2016cvr,CMS:2018zze,CMS:2020geg,CMS:2021nhn} and subtracted separately for \Rsym and \Rasym. In this region, no wake signal is expected due to the $\eta$ ordering requirement $\jtetaone > \jtetatwo$. The resulting \Rsym distribution is then subtracted from \Rasym, isolating the diffusion wake signal~\cite{Yang:2025dqu}. The away-side correlation for large dijet $\eta$ gaps also shows wake-like yield depletion, but the different position of the away-side jet peak in large and small $\eta$ gap samples prevents signal extraction in a data-driven manner.

Various sources of systematic uncertainties are considered. The signal is largely insensitive to jet reconstruction systematic uncertainties, namely, the jet energy scale and JER. In the data, variations of the jet energy scale within its uncertainty and smearing of the jet momentum according to the JER uncertainty produce results consistent with the nominal values, indicating that these effects are negligible. In MC simulations without medium effects, the UE shows no variation with respect to the subleading jet location, as expected, and closure tests (ratios of reconstructed to generated-level distributions) are consistent with unity.

\ifthenelse{\boolean{cms@external}}
{
\begin{figure*}[htbp]
\centering
\includegraphics[width=0.99\textwidth]{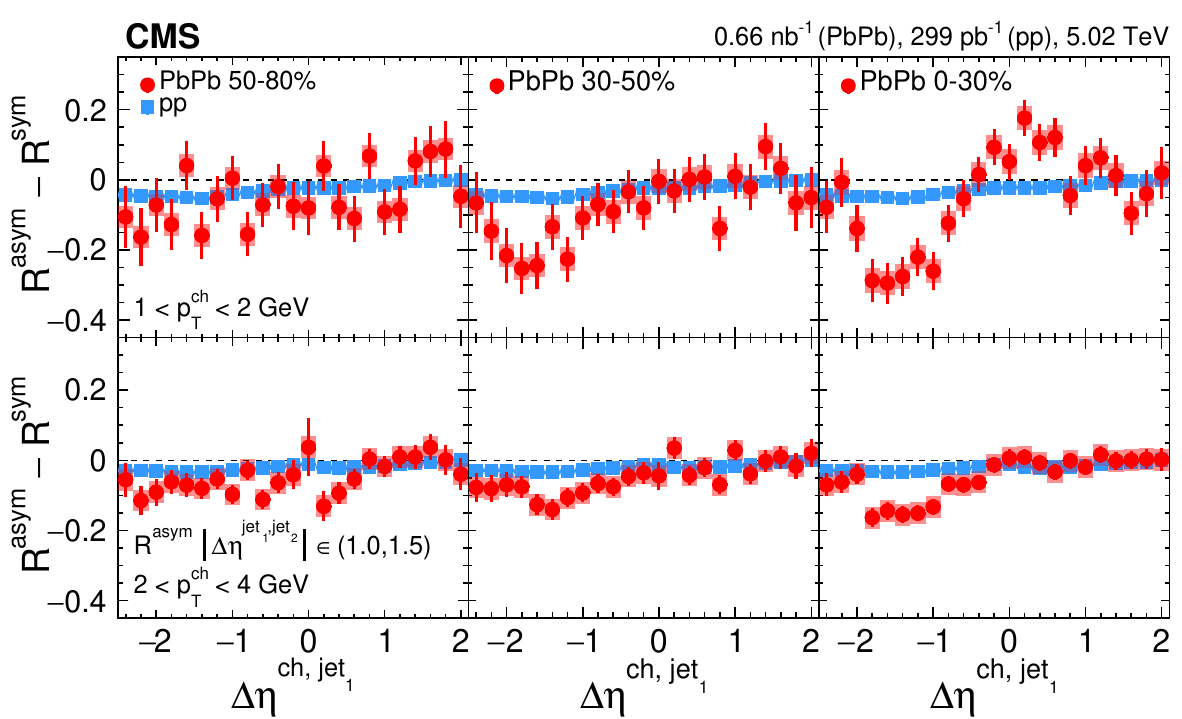}
\caption{The difference of the near-side charged-particle yields $\Rasym-\Rsym$ as a function of \detaldtrk in \pp (blue squares) and different centrality \PbPb collisions (red circles), with the most central (0--30\%) shown in the right panels. Results are shown for two \pth\ ranges: $1 < \pth < 2\GeV$ (upper) and $2 < \pth < 4\GeV$ (lower). Solid vertical lines (shaded areas) show statistical (systematic) uncertainties.}
\label{fig:yield_subtracted}
\end{figure*}
}

The dominant uncertainties arise from the tracking efficiency determination, pair acceptance corrections, and residual UE mismatches between small- and large-gap dijet selections. The tracking efficiency uncertainty is estimated by comparing response to variations in track selections between data and MC simulation~\cite{Khachatryan:2016odn}, resulting in uncertainties of 5\% (2.5\%) for \PbPb (\pp) events~\cite{CMS:2024fli}. The maximum observed variation is conservatively applied to all \detaldtrk bins, corresponding to an absolute uncertainty between 0 and 0.025 across different \pth, centrality, and dijet $\eta$ gap intervals. Pair-acceptance uncertainties are quantified by examining long-range \detaldtrk-dependent asymmetries in the mixed-event-corrected \Rsym distribution, using differences between the negative and positive sidebands ($-2.5 < \detaldtrk < -1.5$ and $1.5 < \detaldtrk < 2.5$), with an absolute uncertainty ranging from 0 to 0.01. Residual UE levels are estimated by comparing the subtracted distributions in the positive \detaldtrk region using different \dphildtrk ranges to perform the projection from the two-dimensional to one-dimensional correlations. The varied ranges ($0 < \dphildtrk < \pi$ and $13\pi/20 < \dphildtrk < 3\pi/2$) are chosen for their sensitivity to long-range correlations. The resulting absolute uncertainty ranges from 0.01 to 0.04. The assigned uncertainty in UE subtraction reflects the level of constraint provided by the data. The effects of the centrality calibration, evaluated by varying the minimum bias event selection efficiency of the HF calorimeters~\cite{CMS:2024krd}, are negligible. The residual UE level mismatch provides the dominant contribution to the systematic uncertainty, followed by the tracking efficiency and pair-acceptance corrections. Uncertainties from all sources are added in quadrature for the total systematic uncertainty.

Figure~\ref{fig:yield_subtracted} shows the near-side correlation differences $\Rasym-\Rsym$ for \pp and \PbPb collisions, with \Rasym having $\detajtonetwo \in (1.0, 1.5)$. In the region of interest, $-2.4 < \detaldtrk < -0.6$, the \pp distributions are approximately flat, with minimal deviation from zero. In contrast, \PbPb collisions show a pronounced dip, most prominent at low \pth. The strength of the depletion increases from peripheral (50--80\%) to central (0--30\%) collisions. The \pp and 50--80\% \PbPb results are statistically consistent, whereas the 0--30\% \PbPb data for $1 < \pth < 2\GeV$ lie significantly below the \pp reference. A similar depletion is expected and is visible on the away-side of \Rasym (as shown in the lower right panel of Fig.~\ref{fig:2Dresults}); however, the displacement of the away-side peaks observed in \Rsym and \Rasym precludes a direct subtraction.

\ifthenelse{\boolean{cms@external}}
{}
{
\begin{figure}[htbp]
\centering
\includegraphics[width=0.99\textwidth]{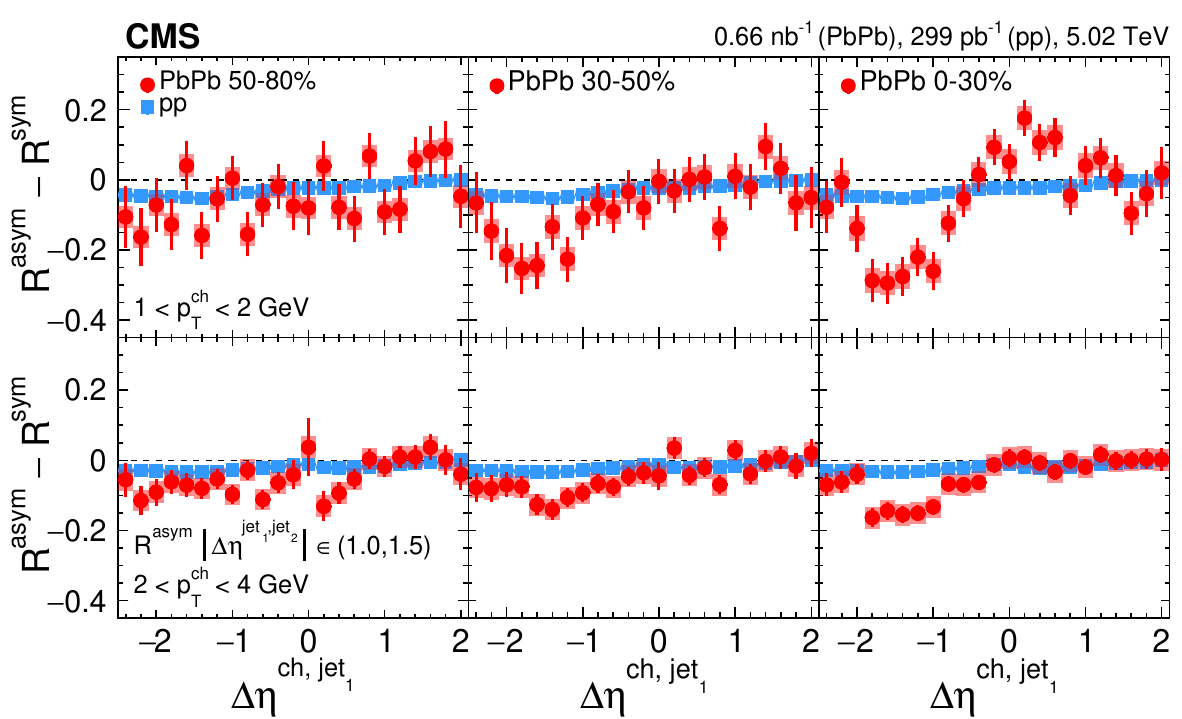}
\caption{The difference of the near-side charged-particle yields $\Rasym-\Rsym$ as a function of \detaldtrk in \pp (blue squares) and different centrality \PbPb collisions (red circles), with the most central (0--30\%) shown in the right panels. Results are shown for two \pth\ ranges: $1 < \pth < 2\GeV$ (upper) and $2 < \pth < 4\GeV$ (lower). Solid vertical lines (shaded areas) show statistical (systematic) uncertainties.}
\label{fig:yield_subtracted}
\end{figure}
}

\begin{figure}[htbp]
\centering
\includegraphics[width=0.49\textwidth]{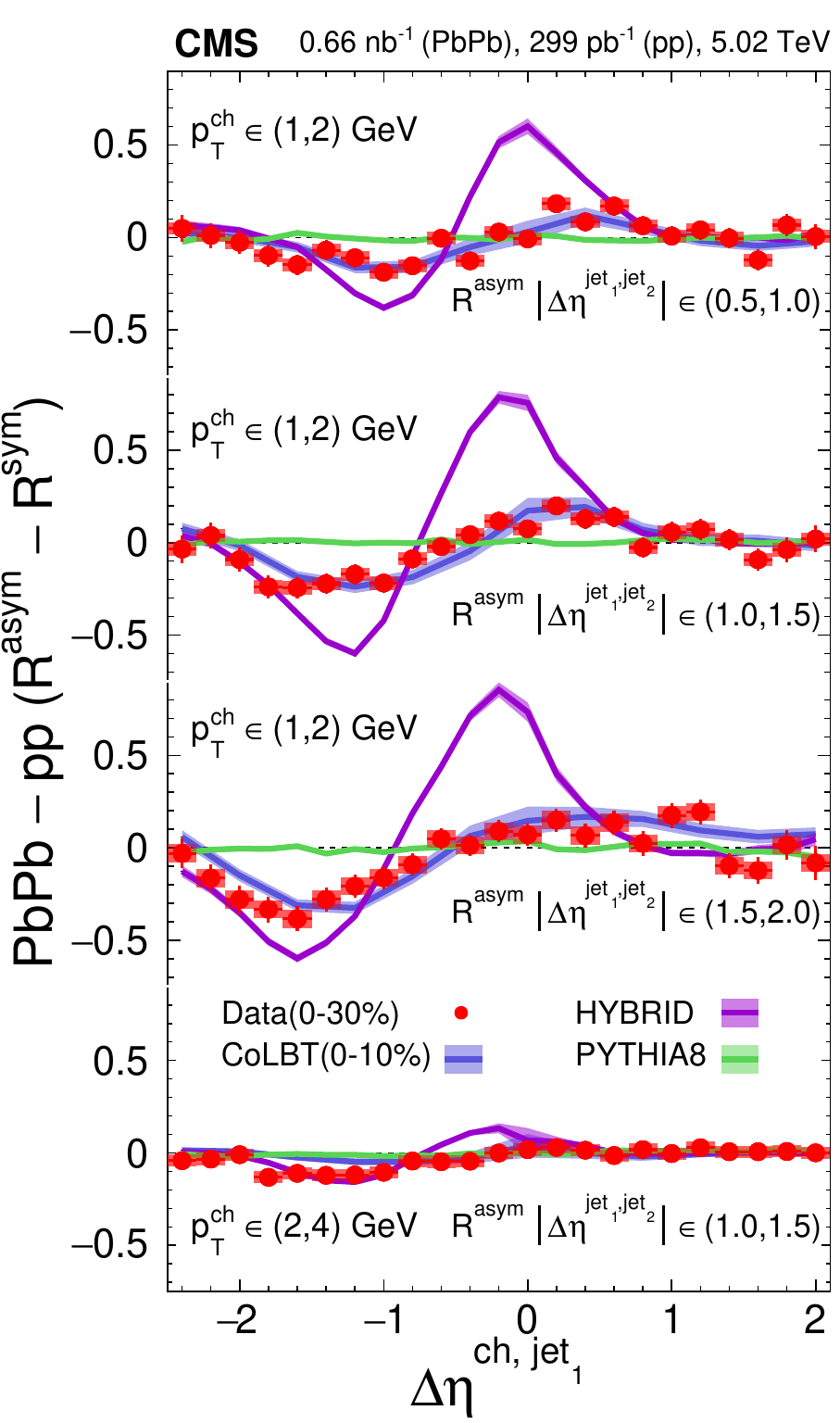}
\caption{The difference between \PbPb and \pp collisions for the particle yield difference observable ($\Rasym-\Rsym$) as a function of \detaldtrk for central (0--30\%) collisions. Results for $1 < \pth < 2\GeV$ are shown for $\Rasym \detajtonetwo \in (0.5, 1.0)$ (upper panel), $\detajtonetwo \in (1.0, 1.5)$ (upper middle panel), and $\detajtonetwo \in (1.5, 2.0)$ (lower middle panel). The result for $2 < \pth < 4\GeV$ with $\Rasym \detajtonetwo \in (1.0, 1.5)$ is shown in the lower panel. Predictions from the \PYTHIA~\cite{pythia82,pythiaCP5tune,nnpdf31,Lokhtin:2005px}, \HYBRID\cite{Casalderrey-Solana:2014bpa, Hulcher:2017cpt}, and CoLBT-hydro~\cite{Chen:2017zte} models are shown as colored bands. Solid vertical lines (shaded areas) show statistical (systematic) uncertainties. For the models, only statistical uncertainties are shown.}
\label{fig:yield_subtracted_Cent030}
\end{figure}

Figure~\ref{fig:yield_subtracted_Cent030} quantifies the medium modification of the yield difference $\Rasym-\Rsym$ by subtracting the \pp data from 0--30\% \PbPb results.  For the $1 < \pth < 2\GeV$ data, the mean position of the observed depletion shifts toward negative \detaldtrk as the \Rasym \detajtonetwo selection increases from $(0.5, 1.0)$ to $(1.5, 2.0)$. Medium-induced jet peak modifications and out-of-cone radiation can interfere with the diffusion wake signal, reducing the strength of the depletion for smaller gaps. As a result, larger \detaldtrk gaps are expected to exhibit stronger diffusion wake signals, which can be seen in Fig.~\ref{fig:yield_subtracted_Cent030}. The results for $2 < \pth < 4\GeV$ with $\Rasym \detajtonetwo \in (1.0, 1.5)$ are shown in the lower panel of the same figure, showing reduced diffusion wake strength consistent with theoretical expectations.

Figure~\ref{fig:yield_subtracted_Cent030} also presents comparisons with theoretical models. The \PYTHIA model with the CP5 tune (green bands) shows no $\detaldtrk$ dependence, as expected in the absence of jet quenching. The \HYBRID model, combining DGLAP-based weakly coupled jet evolution with strongly coupled interactions between jet constituents and the QGP~\cite{Casalderrey-Solana:2014bpa}, is shown as purple bands for the fully incoherent energy loss scenario~\cite{Hulcher:2017cpt}. The model overpredicts the observed depletion for $1 < \pth < 2\GeV$, while better capturing the data trends for $2 < \pth < 4\GeV$. The enhancement near $\detaldtrk \approx 0$ arises in the model from the diffusion wake caused by the subleading jet that should reduce the near-side peak magnitude in \Rsym. A much smaller enhancement than predicted by the HYBRID model is observed in 0--30\% central data for $1 < \pth < 2\GeV$. The CoLBT-hydro model (blue bands) describes jet propagation and jet-induced medium excitation by coupling linear Boltzmann transport for parton energy loss with a (3+1)D hydrodynamic evolution of the QGP~\cite{Chen:2017zte}. For this model, predictions are available for 0--10\% central collisions. Compared with \HYBRID, the model shows better agreement with the data for both negative and positive yields, although a direct comparison is impaired by differences in centrality. The mean \detaldtrk position of the negative dip appears to be captured by both the \HYBRID and CoLBT models.

\begin{figure}[htbp]
\centering
\includegraphics[width=0.49\textwidth]{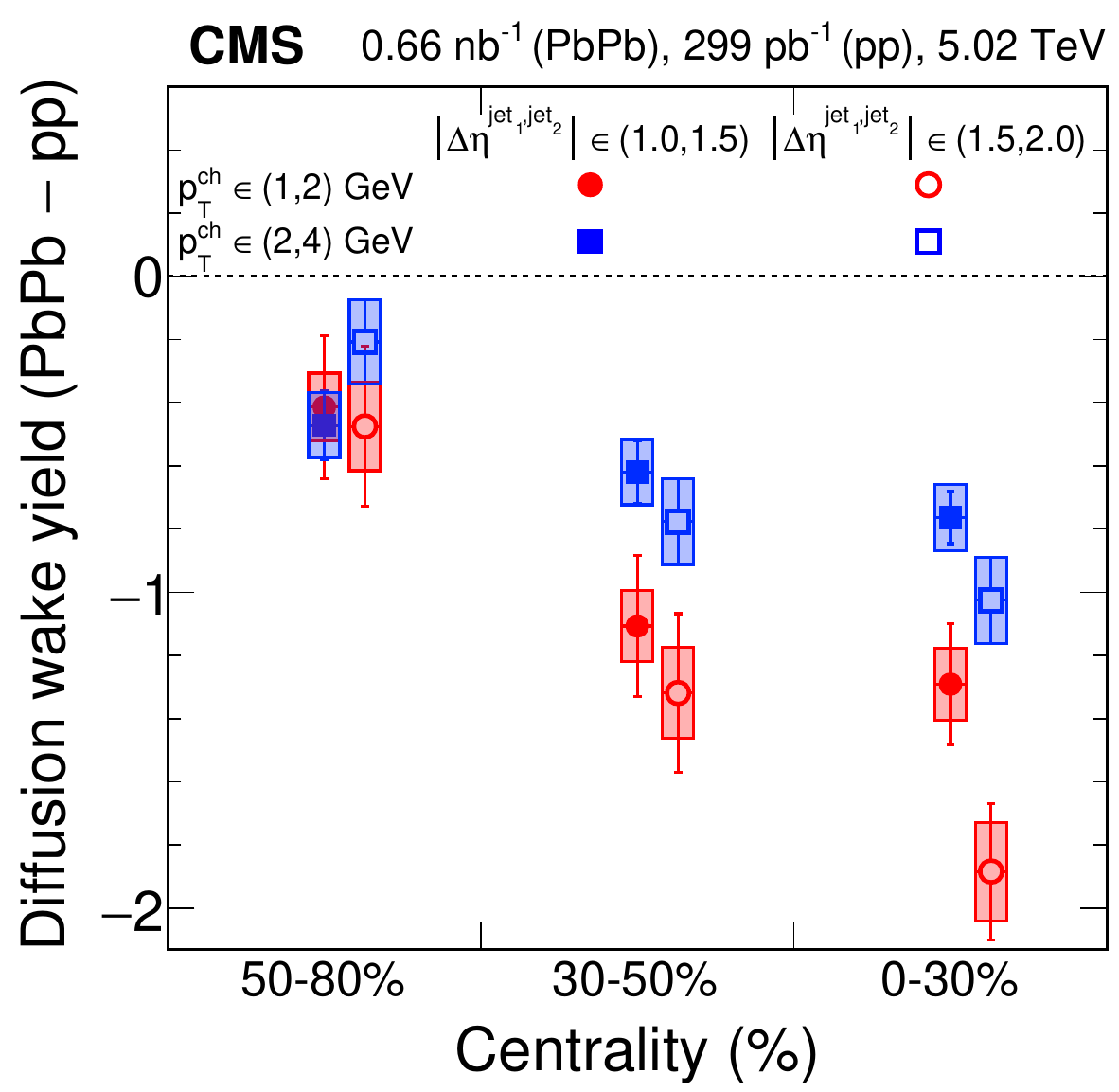}
\caption{Charged-particle yield difference integrated over $-2.4 < \detaldtrk < -0.6$ as a function of \PbPb collision centrality for $\detajtonetwo \in (1.0, 1.5)$ (closed markers) and $\detajtonetwo \in (1.5, 2.0)$ (open markers). Red markers indicate $1 < \pth < 2\GeV$, and blue markers indicate $2 < \pth < 4\GeV$. Solid vertical lines (shaded areas) show statistical (systematic) uncertainties. The 0--30\% data deviate from zero by more than 5 standard deviations from a $\chi^2$ test.}
\label{fig:wake_integral}
\end{figure}

Figure~\ref{fig:wake_integral} quantifies the centrality and \pth dependences of the diffusion wake signal using medium-modified correlated yields integrated over $-2.4 < \detaldtrk < -0.6$ for $\Rasym \detajtonetwo \in (1.0, 1.5)$ and $\Rasym \detajtonetwo \in (1.5, 2.0)$. The depletion of the yields is greatest in 0--30\% collisions, with the $1 < \pth < 2\GeV$ data deviating from zero by more than 5 standard deviations for both \detajtonetwo selections based on a $\chi^2$ test. The yield depletion is smaller for more peripheral collisions, higher \pth selections, and smaller dijet $\eta$ separations. The diffusion wake effect in the QGP is the only known mechanism capable of producing such structures in the UE distribution in the direction opposite to the propagating jet, and therefore, this work provides an observation of the diffusion wake effect in the QGP.

In summary, we report the first observation of the diffusion wake using dijets in heavy ion collisions. The measurements use pseudorapidity-separated dijets in lead-lead (\PbPb) and proton-proton (\pp) collisions at a center-of-mass energy per nucleon pair of 5.02\TeV. Charged-particle correlations were studied relative to the highest transverse momentum (\pt) jet of each dijet. Correlations are measured for dijets with different pseudorapidity separations, and the difference between correlations from large and small separation samples is used for the jet diffusion wake extraction. When isolating medium modifications by examining the difference between \PbPb and \pp collisions, a pronounced depletion of charged-particle yields is observed in the region where a wake signal is expected. The integrated signal for low-\pt charged particles ($1 < \pth < 2\GeV$) in central collisions deviates from zero with a significance exceeding 5 standard deviations. The depletion is less pronounced in more peripheral collisions and when examining a higher \pth selection. Theoretical models incorporating jet energy loss and medium response show qualitatively similar trends with those observed in the data, whereas models without jet-medium interactions do not show any depletion. This observation of the diffusion wake offers new insights into quark-gluon plasma properties and jet-induced medium response.

\begin{acknowledgments}
We congratulate our colleagues in the CERN accelerator departments for the excellent performance of the LHC and thank the technical and administrative staffs at CERN and at other CMS institutes for their contributions to the success of the CMS effort. In addition, we gratefully acknowledge the computing centers and personnel of the Worldwide LHC Computing Grid and other centers for delivering so effectively the computing infrastructure essential to our analyses. Finally, we acknowledge the enduring support for the construction and operation of the LHC, the CMS detector, and the supporting computing infrastructure provided by the following funding agencies: SC (Armenia), BMBWF and FWF (Austria); FNRS and FWO (Belgium); CNPq, CAPES, FAPERJ, FAPERGS, and FAPESP (Brazil); MES and BNSF (Bulgaria); CERN; CAS, MoST, and NSFC (China); MINCIENCIAS (Colombia); MSES and CSF (Croatia); RIF (Cyprus); SENESCYT (Ecuador); ERC PRG and PSG, TARISTU24-TK10 and MoER TK202 (Estonia); Academy of Finland, MEC, and HIP (Finland); CEA and CNRS/IN2P3 (France); SRNSF (Georgia); BMFTR, DFG, and HGF (Germany); GSRI (Greece); NKFIH (Hungary); DAE and DST (India); IPM (Iran); SFI (Ireland); INFN (Italy); MSIT and NRF (Republic of Korea); MES (Latvia); LMTLT (Lithuania); MOE and UM (Malaysia); BUAP, CINVESTAV, CONACYT, LNS, SEP, and UASLP-FAI (Mexico); MOS (Montenegro); MBIE (New Zealand); PAEC (Pakistan); MES, NSC, and NAWA (Poland); FCT (Portugal);  MESTD (Serbia); MICIU/AEI and PCTI (Spain); MOSTR (Sri Lanka); Swiss Funding Agencies (Switzerland); MST (Taipei); MHESI (Thailand); TUBITAK and TENMAK (T\"{u}rkiye); NASU (Ukraine); STFC (United Kingdom); DOE and NSF (USA).
\end{acknowledgments}
\bibliography{auto_generated}

\clearpage

\begin{figure*}[htbp!]
\centering
\includegraphics[width=0.49\textwidth]{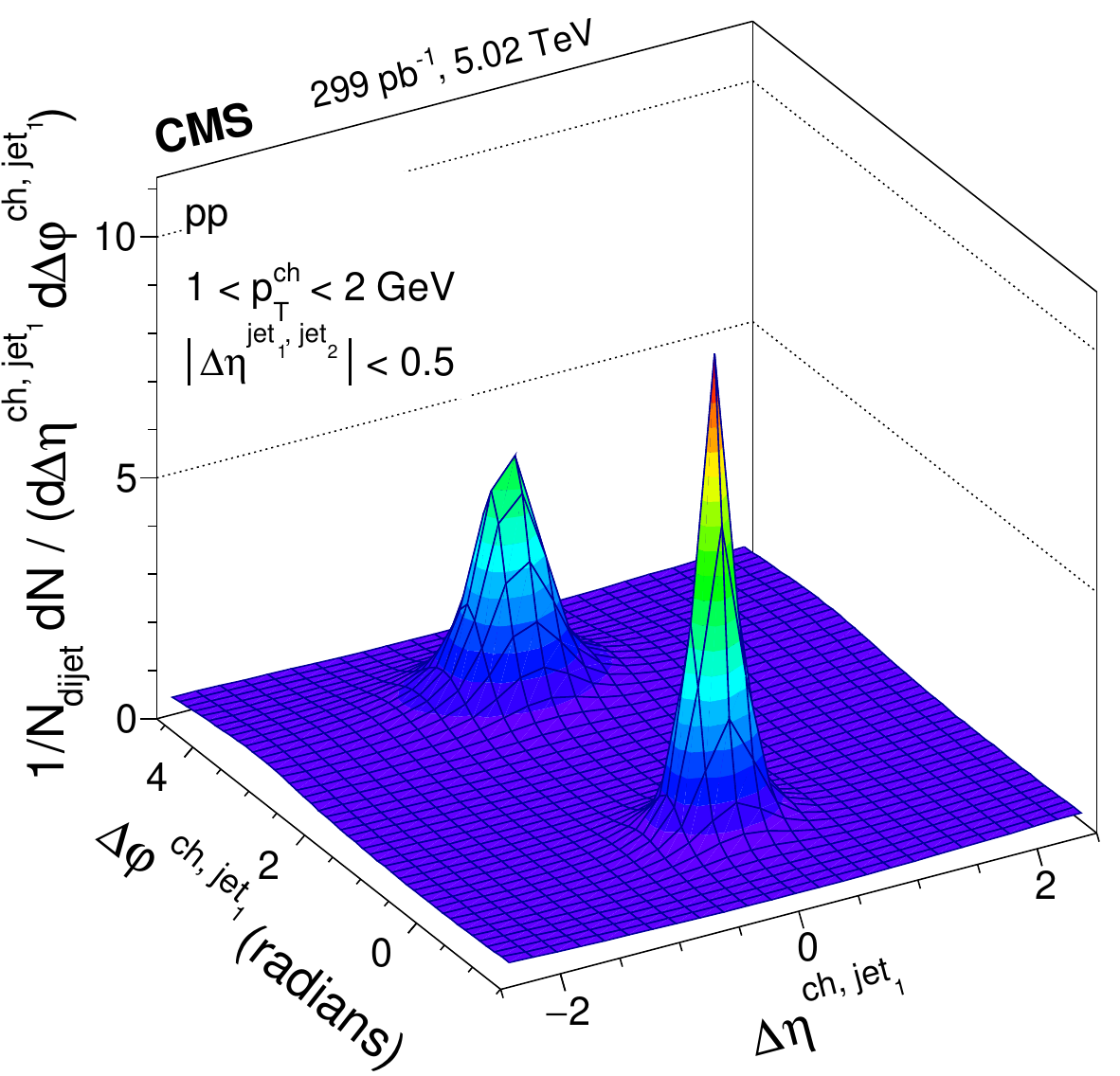}
\includegraphics[width=0.49\textwidth]{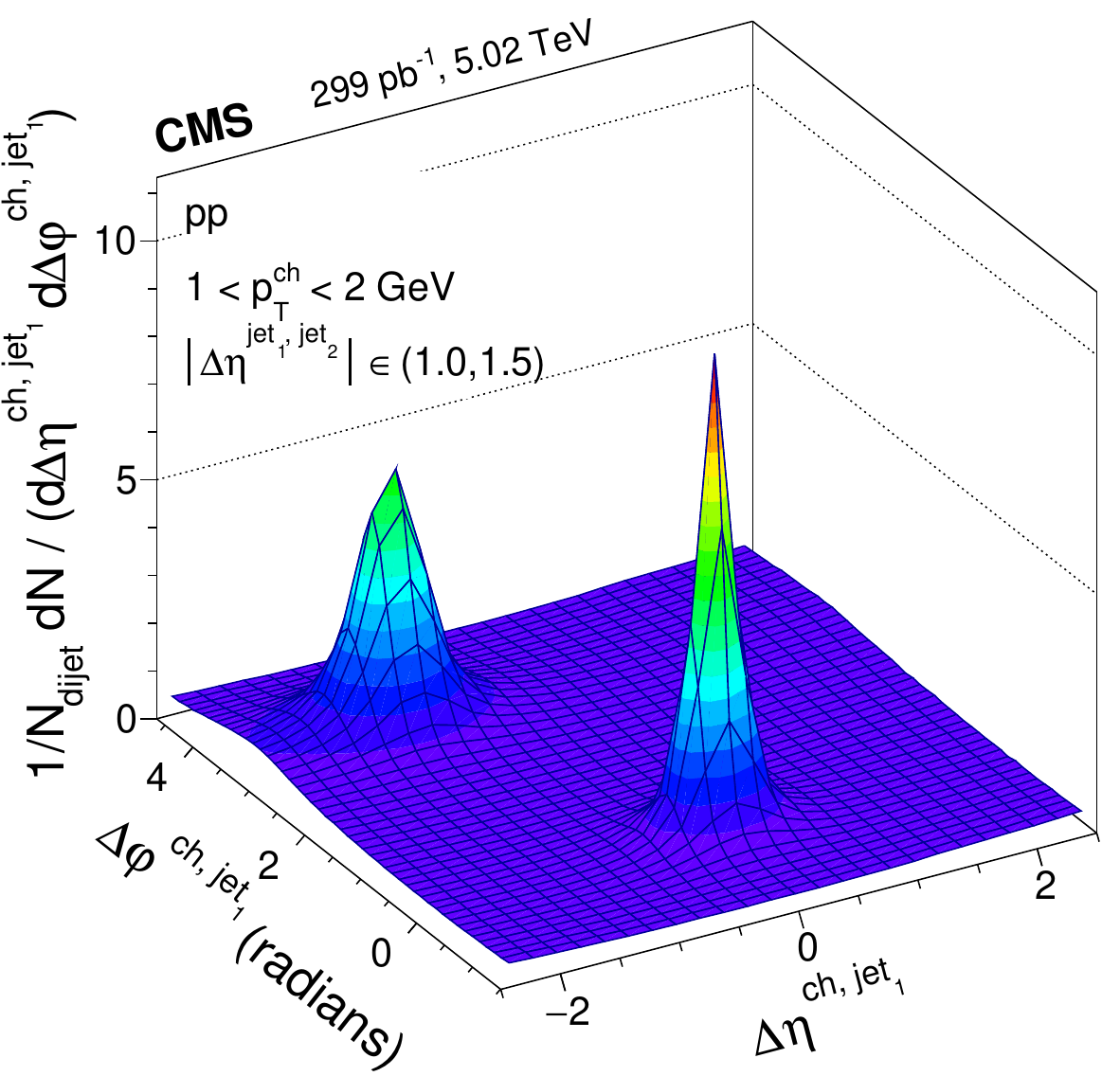}
\includegraphics[width=0.49\textwidth]{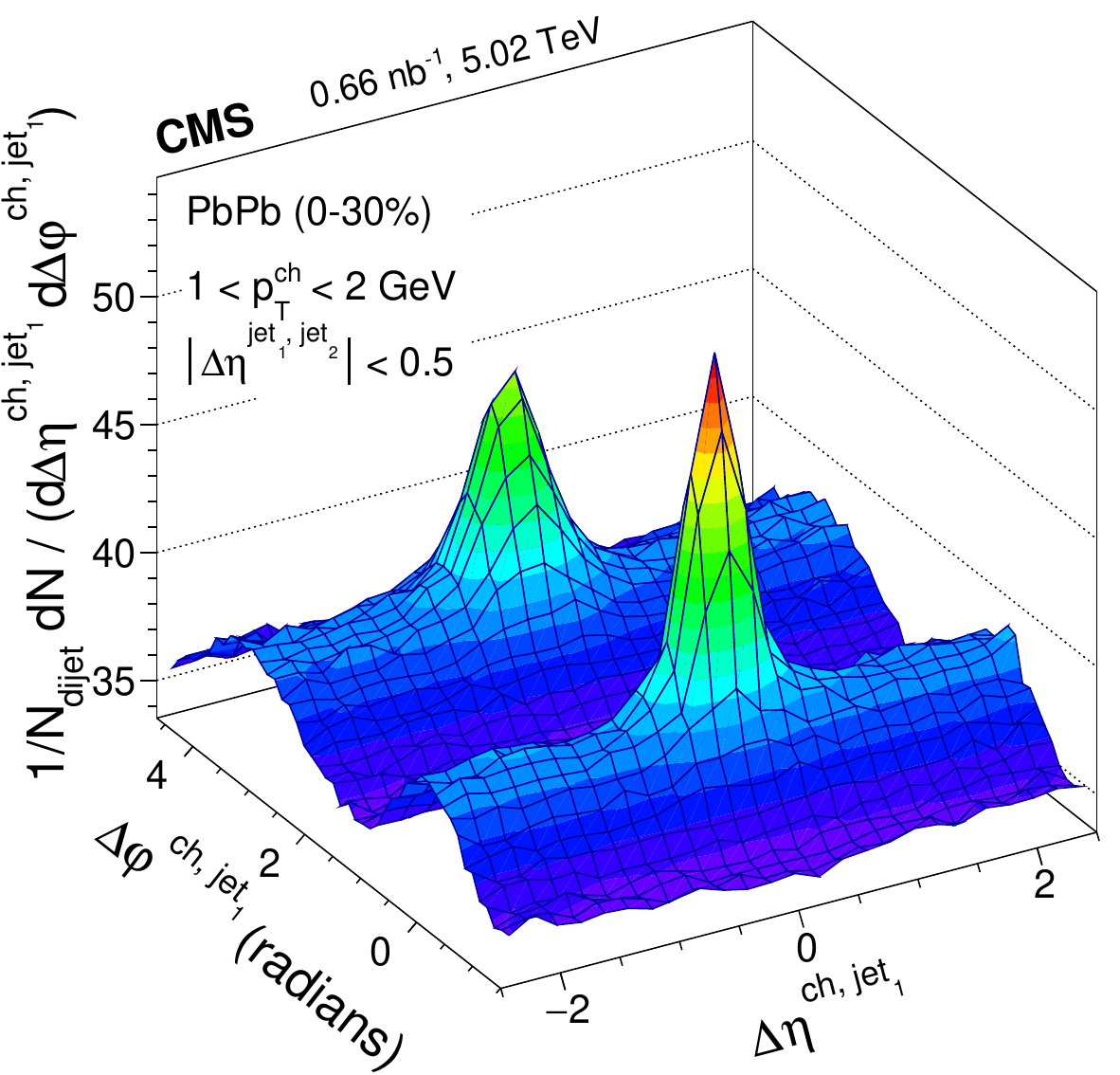}
\includegraphics[width=0.49\textwidth]{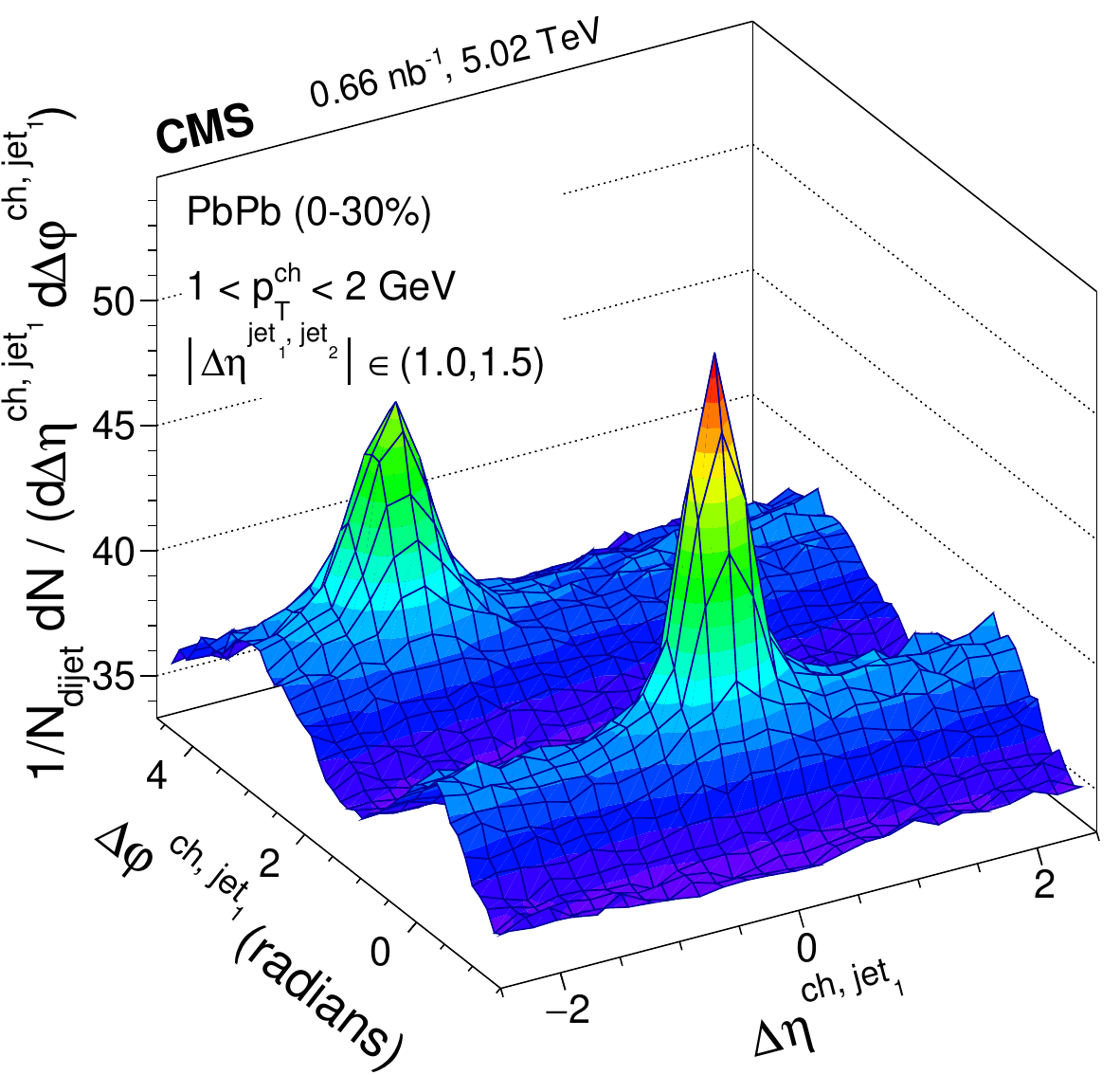}
\caption{Distributions of charged-particle yields with $1 < \pth < 2\GeV$ as functions of $\detaldtrk$ and $\dphildtrk$, measured relative to the leading jet direction in \pp\ (upper panels) and 0--30\% \PbPb\ (lower panels) collisions. The left (right) column shows the results for small $(\Rsym\,\detajtonetwo < 0.5)$ and large $[\Rasym\,\detajtonetwo \in (1.0,1.5)]$ dijet configurations.}
\label{fig:2Dresults}
\end{figure*}

\appendix
\numberwithin{figure}{section}
\ifthenelse{\boolean{cms@external}}{\section{Two-dimensional angular correlations}\label{app:2D_jetTrackCorr}}{\section{Two-dimensional angular correlations}\label{app:2D_jetTrackCorr}}
Figure~\ref{fig:2Dresults} shows the two-dimensional distributions of charged-particle yields in \pp and 0--30\% \PbPb collisions for $1<\pth<2\GeV$, as functions of $\detaldtrk$ and $\dphildtrk$ relative to the leading jet.  Distributions for small $(\Rsym\,\detajtonetwo < 0.5)$ and large $[\Rasym\,\detajtonetwo \in (1.0,1.5)]$ dijet $\eta$ gaps are shown. Charged-particle tracks associated with the fragmentation of the leading jet produce a peak near $(\detaldtrk,\dphildtrk) \approx (0,0)$, while those from the subleading jet cluster around $(\detaldtrk,\dphildtrk) \approx (0,\pi)$ or $(-1,\pi)$ for small and large $\eta$ gaps, respectively. These distributions are then projected onto $\detaldtrk$ to isolate the wake signal.

\cleardoublepage \section{The CMS Collaboration \label{app:collab}}\begin{sloppypar}\hyphenpenalty=5000\widowpenalty=500\clubpenalty=5000\cmsinstitute{Yerevan Physics Institute, Yerevan, Armenia}
{\tolerance=6000
A.~Hayrapetyan, V.~Makarenko\cmsorcid{0000-0002-8406-8605}, A.~Tumasyan\cmsAuthorMark{1}\cmsorcid{0009-0000-0684-6742}
\par}
\cmsinstitute{Institut f\"{u}r Hochenergiephysik, Vienna, Austria}
{\tolerance=6000
W.~Adam\cmsorcid{0000-0001-9099-4341}, L.~Benato\cmsorcid{0000-0001-5135-7489}, T.~Bergauer\cmsorcid{0000-0002-5786-0293}, M.~Dragicevic\cmsorcid{0000-0003-1967-6783}, P.S.~Hussain\cmsorcid{0000-0002-4825-5278}, M.~Jeitler\cmsAuthorMark{2}\cmsorcid{0000-0002-5141-9560}, N.~Krammer\cmsorcid{0000-0002-0548-0985}, A.~Li\cmsorcid{0000-0002-4547-116X}, D.~Liko\cmsorcid{0000-0002-3380-473X}, M.~Matthewman, J.~Schieck\cmsAuthorMark{2}\cmsorcid{0000-0002-1058-8093}, R.~Sch\"{o}fbeck\cmsAuthorMark{2}\cmsorcid{0000-0002-2332-8784}, M.~Shooshtari\cmsorcid{0009-0004-8882-4887}, M.~Sonawane\cmsorcid{0000-0003-0510-7010}, N.~Van~Den~Bossche\cmsorcid{0000-0003-2973-4991}, W.~Waltenberger\cmsorcid{0000-0002-6215-7228}, C.-E.~Wulz\cmsAuthorMark{2}\cmsorcid{0000-0001-9226-5812}
% ...
\par}
\vskip\cmsinstskip
\dag:~Deceased\\
$^{1}$Also at Yerevan State University, Yerevan, Armenia\\
$^{2}$Also at TU Wien, Vienna, Austria\\
% ...
\end{sloppypar}
\end{document}